# Synthesis of novel rare earth – iron oxide chalcogenides with the $La_2Fe_2O_3Se_2$ structure

**D.O. Charkin*[1], V.A. Plotnikov[1], A.V.Sadakov[2], O.E.Omel'yanovskii[2], and S.M. Kazakov[1]**

[1] *Department of Chemistry, Moscow State University, Leninskie Gory 1, Moscow 119991 Russia*

[2] *P.N. Lebedev Physical Institute, RAS, Moscow, 119991, Russia*

**Abstract**

Our searches for new oxide chalcogenides of rare earths and Fe, Co, Ni, and Zn resulted in preparation of two new compounds $Ce_2Fe_2O_3S_2$ and $Pr_2Fe_2O_3S_2$ which are isostructural to $La_2Fe_2O_3Se_2$ and $Sm_2Ti_2O_3Sb_2$. Crystal structures of the new compounds $Ce_2Fe_2O_3S_2$ and $Pr_2Fe_2O_3S_2$ were determined from powder X-ray diffraction data. Magnetic measurements were performed for $Ce_2Fe_2O_3S_2$ and revealed behavior very similar to that of isostructural oxide chalcogenides of iron and pnictides of titanium. In particular, no superconductivity was observed down to 4 K. Crystal chemical factors determining the stability of the $La_2Fe_2O_3Se_2$ structure type are discussed.

**Keywords:** A. Semiconductors; B. Solid state reactions; C. Crystal structure; D. Magnetic measurements; D. X-ray diffraction

## 1. Introduction

Oxide chalcogenides and pnictides of *f*-elements, particularly of rare earths, exhibit a diversity of non-trivial crystal structures, of which a large part is layered [1]. These compounds have attracted attention first as potential fast ionic conductors [2], next, as transparent semiconductors, and recently, as medium-temperature superconductors (for a review on the last two issues, as well as general crystal chemistry, see [1] and references therein). The superconducting properties are now being associated with oxide pnictides of rare earths and iron [3, 4], as well as structurally related binary iron chalcogenides [5]. There exist a few oxide chalcogenides of rare earths and iron as well as cobalt or manganese [6 – 9] with a general formula $Ln_2M_2O_3Ch_2$ (M = Mn, Fe, or Co, and Ch = S or Se) which share a common structure type somewhat different from those of quaternary "1111" Ln – Fe (or Ln – Co) oxide pnictides, as well as binary Fe chalcogenides (Fig. 1). The peculiarity of the structure is in transition metal residing in a multi-ligand environment, i.e. a *trans*-$O_2Ch_4$ octahedron

* Corresponding author. Email: charkin@inorg.chem.msu.ru (D.O. Charkin)

(Ch = S or Se). A small group of alkaline earth compounds containing $[Ae_2F_2]^{2+}$ layers (Ae = Sr or Ba) instead of $[Ln_2O_2]^{2+}$ is also known [10]. Studies of magnetic properties of these compounds were undertaken but revealed no superconducting material as yet.

When considering multinary rare earth compounds, two questions are often addressed, (i) how far an isostructural series may be stretched upon varying the nature of a $Ln^{3+}$ cation and keeping otherwise the chemical composition intact, and (ii) whether it is possible, for the same rare earth cation, to replace the large anion (chalcogen or pnicogen) by its heavier analogs within the same structure type. A semi-quantitative analysis had been done by us for a series of Ln – Cu and Ln – Ag oxide chalcogenides with LaOAgS structure [11]. In the current paper, we report the results of our searches for new representatives of the $La_2Fe_2O_3Se_2$ structure type among oxide chalcogenides of Ln and Fe, Co, Ni, and Zn, and attempt to compare the tendencies with those found earlier for the LaOAgS structure type.

## 2. Experimental

### 2.1. Synthesis

The starting materials were metallic rare earths (La – Sm), iron, cobalt, nickel, and zinc, oxides $Fe_2O_3$, $CeO_2$, and $Pr_2O_3$, sulfur, selenium, and tellurium. All chemicals were of analytical or extra purity grade. The initial charges were ground, pressed into pellets, sealed in evacuated silica tubes, slowly heated to the target temperature, re-ground, re-pressed, and re-annealed. Exact synthesis conditions and results of primary X-ray analysis are collected in Table 1.

### 2.2. Magnetic measurements

Magnetic susceptibility measurement was performed using a Quantum Design PPMS ac-susceptometer. In this standard method an alternating magnetic field is applied to the sample via copper drive coil, and a detection coil set (two counterwound copper coils connected in series) inductively responds to the combined sample moment and excitation field. The sample was placed inside one of the detection coils. Amplitude and frequency of the applied ac-field were 1 Oe and 137 Hz, respectively. The samples were cooled down in zero dc-field.

## 3. Results and discussion

### 3.1. New compounds and crystal structures

As follows from Table 1, four compounds were found in our studies, including two new oxysulfides $Ce_2Fe_2O_3S_2$ and $Pr_2Fe_2O_3S_2$. The new oxysulfides are black powders stable in air for long periods of time. They could not be prepared completely free from minor Ln–O–Ch or Ln–O–M admixtures, and the purity of sulfide samples was higher. Therefore, structural parameters have been determined for $Ce_2Fe_2O_3S_2$ and $Pr_2Fe_2O_3S_2$ using powder data. The small-level impurities found in the samples were included in the refinement. Powder X-ray diffraction data were collected on a Stoe ($Ce_2Fe_2O_3S_2$) and Bruker D8 Advance ($Pr_2Fe_2O_3S_2$) diffractometers ($CuK\alpha_1$ radiation, Ge(111) monochromator, reflection geometry) Rietveld refinements were performed with the TOPAS package [12] using the fundamental parameter approach as reflection profiles. The structural model proposed by Mayer *et al* [6] was used for the starting coordinate values. Preferred orientation was corrected using a spherical harmonics approach developed in TOPAS. Further experimental details, refined atomic parameters, and main structural parameters are collected in Tables 2 – 4. Final Rietveld refinement plot for $Ce_2Fe_2O_3S_2$ is shown on Fig. 2.

Upon comparison of the structures for $Ce_2Fe_2O_3S_2$ and $Pr_2Fe_2O_3S_2$, one can easily see that the cell volume and parameters of the Ce compound is smaller than that of Pr, a common phenomenon in rare earth oxide chalcogenide or pnictide series, most pronounced for coinage metal compounds LnOMCh (Ch = S, Se, or Te, and M = Cu or Ag). The “cerium pit” is geometrically reflected by abnormally short Ce – Ch distances and usually accompanied by a large thermal parameter of the transition metal. The actual reason of this phenomenon is not completely established. One suggestion is that there is a partial oxidation of $Ce^{III}$ to $Ce^{IV}$ and partial reduction of the coinage metal with generation of vacancies at its site, and it is illustrated by preparation of non-stoichiometric $CeOAg_{1-x}Se$ [13], $CeOCu_{1-x}Se$, and $CeOCu_{1-x}Te$ [14]. On the other hand, stoichiometric CeOCuS can also be obtained and contains only trivalent cerium [15], and anomalously short Ce – S distance was explained in terms of Ce – S bond covalence and hybridization of Ce orbitals. It is interesting to note that the “cerium pit” is not observed in the $Ln_2Fe_2O_3Se_2$ series. We also point that it is not observed for LnOMnPn and $Ln_2Mn_2O_3Se_2$ series, and $Mn^{2+}$ is suggested to be more electropositive than $Fe^{2+}$ (for instance, there exist $LnFe_2Pn_2$ or $AnFe_2Pn_2$ compounds with formal oxidation state of iron below +2 while analogous Mn compounds are unknown).

### 3.2. Magnetic properties

Temperature dependences of magnetic susceptibility of $Ce_2Fe_2O_3S_2$ is presented in Fig. 3. The inset is a $1/\chi$ (T) plot, which is useful to check whether the susceptibility obeys the Curie-Weiss law. The $\chi$=C/(T-$\theta$) behavior is observed above 130K, with $\theta$<0 characteristic for *anti*-ferromagnetics. A more complex behavior below 130K, including a feature around 70K is yet to be clarified. The $\theta$ value (-270K) may have the same order of magnitude as $T_N$ which, for analogous compounds, usually lies within 170 – 180K.

### 3.3. Crystal chemical remarks

Our investigations have established the “terminations points” of the isostructural series $Ln_2Fe_2O_3S_2$ and $Ln_2Co_2O_3Se_2$ to lie at Ln = Pr and La, respectively; the $Ln_2Fe_2O_3Se_2$ and $Ln_2Mn_2O_3Se_2$ series terminate at Ln = Sm and Pr. No isostructural nickel or tellurium compounds have been found. The three isostructural series appear very short as compared to LnOTPn (T = Mn, Fe or Co, and Pn = P or As) and the only common observation is that no compound has been prepared yet involving large $Te^{2-}$/$Sb^{3-}$ anions. The early termination of the $Ln_2T_2O_3Ch_2$ series as compared to LnOTPn may be explained in terms of distortion of the $TO_2Ch_4$ octahedra: for instance, given the Fe–S distance in $Ce_2Fe_2O_3S_2$ of 2.633(3)Å, the size of the S–S edge of a regular $FeS_4$ square is expected to be 3.72Å which is much smaller than the edge of the actual distorted octahedron which equals the *a* cell parameter of 4.00Å. For $Sm_2Fe_2O_3Se_2$, the Se–Se edge of the distorted octahedron is 4.00Å [9] which is again larger than the "ideal" value of 3.82Å calculated from Fe–Se bond distance of 2.71Å [7]. This distortion would indeed decrease when smaller $Ln^{3+}$ are involved, but there is another reason of instability, the somewhat compressed T–O bond which equals one half of the cell parameter. Note that the *a* cell parameters, thus the Fe–O bond lengths, almost coincide for terminating points of both iron-containing series. These values lie well below the cell parameters of the $Ae_2Fe_2OF_2Ch_2$ compounds described in [10]. The stability area of the $La_2Fe_2O_3Se_2$ structure type reflects a delicate interplay involving two kinds of distortions (T–Ch bond distances and Ch–T–Ch bond angles) in the anionic layers. In $La_2Co_2O_3Se_2$, the $CoSe_4$ squares are even more distorted as compared to $La_2Fe_2O_3Se_2$ and even $Nd_2Fe_2O_3Se_2$ (in terms of Ch–T–Ch angles). It is thus evident that the $Ln_2Co_2O_3Se_2$ series should terminate earlier than $Ln_2Fe_2O_3Se_2$, and explains why nickel compounds, as well as cobalt sulfides, were not observed. We also note that no oxide chalcogenide was found for zinc while there exists a cadmium compound, $La_2CdO_2Se_2$ [16] with a LaOAgS-related structure.

When turning to tellurides, the distortions may also appear intolerable as yet another factor comes into play: the large cell parameters observed for rare-earth oxide tellurides suggest also significant distortions of $OLn_4$ tetrahedra of the $[Ln_2O_2]$ layers [11]. This problem is easily overcome using fluorite layers composed of larger $Ae^{2+}$ and $F^-$ ions. In fact, there exist three isostructural oxide pnictides with titanium including two antimony compounds, $Sr_2Ti_2F_2OPn_2$ (Pn = As and Sb) and $Sm_2Ti_2O_3Sb_2$ [17]. In the arsenide, the distortions of the $TiAs_4$ square are very close to those of $La_2Co_2O_3Se_2$ but the $FSr_4$ tetrahedra are perfectly regular. In the antimonides, the pattern is reverse and now the $TiO_2Sb_4$ octahedra are almost undistorted, due to favourable ratio of Ti–Sb and Ti–O bonds. The same applies also to $Ln_2Mn_2O_3Se_2$ series where existence of a few members with later rare-earths is possible. We may suggest that among alkaline earth fluoride compounds, the yet missing $[Fe_2OTe_2]^{2-}$, $[Co_2OS_2]^{2-}$, and $[Ni_2OSe_2]^{2-}$ layers are more likely to be stabilized. It is possible that compounds of nickel, as well as later rare earths, may be prepared using high pressure technique. Formation of $Ln_2Mn_2O_3S_2$ is also likely. By analogy with $Na_2Ti_2OPn_2$ (Pn = As, Sb) [18], we suggest existence of isostructural oxide chalcogenides like $Na_2Mn_2OCh_2$, $Na_2Fe_2OCh_2$ or $Na_2Co_2OCh_2$.

### 3.3. Magnetic properties

In $Ce_2Fe_2O_3S_2$, there is a possibility of charge doping to the $[Fe_2OS_2]^{2-}$ layers by a reducing agent like $Ce^{3+}$ which might induce superconductivity by analogy to that observed for the iron oxide pnictides. The observed pattern (Fig. 3) is rather similar to those of $La_2Fe_2O_3Se_2$ [7] and $Sr_2Ti_2F_2OPn_2$ (Pn = As or Sb) [17] but different from those registered for the Mn compounds [9]. No superconductivity is observed akin to all isostructural compounds. It looks thus likely that superconductivity is favored only upon tetrahedral coordination of $Fe^{2+}$ by $Ch^{2-}$ (Ch = S or Se), since the combinations of Fe and Ch orbitals in layers comprised of $FeCh_4$ tetrahedra and of $FeO_2Ch_4$ octahedra are evidently quite different. The known Ti – As or Ti – Sb oxyphictides of this structure type also do not exhibit superconducting properties. Observation of a CDW in $Na_2Ti_2OPn_2$ was considered as a hint to possible superconductivity in other compounds involving $[Ti_2OPn_2]^{2-}$ layers, but apparently (yet) not in the (undoped) $La_2Fe_2O_3Se_2$ structure [17, 18]. It would be evidently of interest to look if compounds containing $[Fe_2OPn_2]^{4-}$ layers can exist (which may be combined, for instance, with $[Th_2O_2]^{4+}$ or $[Ln_2F_2]^{4+}$ slabs) and whether these compounds exhibit any interesting magnetic properties.

### 4. Conclusions

To summarize, the family of $La_2Fe_2O_3Se_2$-type rare-earth oxide chalcogenides has been extended by four new iron-containing members. Commonly to other oxide chalcogenides, the sulfides are prepared easier and in purer form. The family of isostructural cobalt compounds could not as yet be extended further, and Ni, Zn, and Ru seem not to participate in this family at all. The relatively narrow “crystal chemical boundaries” of this structure type can be explained in terns of two oppositely directed deformations of the *trans*-$TO_2Ch_4$ octahedra (T – transition metal, Ch – chalcogen) condensed into the $[T_2OCh_2]^{2-}$ slabs, the stretching or compression of the axial T–O bonds and rectangular distortion of the equatorial $TCh_4$ squares. As yet, neither representative of the $La_2Fe_2O_3Se_2$-type compounds was found to exhibit superconducting properties typical for the “11” (FeS), “111” (CeFeSi), “122” ($ThCr_2Si_2$), or “1111” (LaOAgS) type iron chalcogenides or pnictides which can be due to non-tetranedral and/or mixed-ligand environment of transition metal (particularly iron) in the $La_2Fe_2O_3Se_2$ structure.

### Acknowledgements

This work was partially supported by the Ministry of Science and Education of Russian Federation under the State contract P-279. The support of the Russian Foundation for Basic Research is acknowledged (Grants No. 10-03-00681-a, 10-02-01281). A.V.S. would like to thank LPI Educational Research Center for the support.

Table 1. Studied compositions and cell parameters of the obtained compounds

| Target composition | Synthesis procedure | X-ray results | Cell parameters, Å |
|---|---|---|---|
| $Ce_2Fe_2O_3S_2$ | $2Ln + 2Ch + Fe_2O_3$ Slow heating to 600°C, annealing for 2 days, regringing and annealing at 1000°C | $Ce_2Fe_2O_3S_2$, FeS | $a = 4.0117(2)$, $c = 17.6888(9)$ |
| $Pr_2Fe_2O_3S_2$ | | $Pr_2Fe_2O_3S_2$ | $a = 4.0093(1)$, $c = 17.6639(4)$ |
| $Ce_2Fe_2O_3Se_2$ | | $Ce_2Fe_2O_3Se_2$ | $a = 4.0623(1)$, $c = 18.4999(9)$ |
| $Pr_2Fe_2O_3Se_2$ | | $Pr_2Fe_2O_3Se_2$ | $a = 4.0473(1)$, $c = 18.4471(9)$ |
| $Nd_2Fe_2O_3Se_2$ | $Ln_2O_3 + 2FeSe$, the same scheme as above | $Nd_2Fe_2O_3Se_2$ | $a = 4.0302(2)$, $c = 18.441(1)$ [a)] |
| $Sm_2Fe_2O_3Se_2$ | | $Sm_2O_2Se$, FeO, FeSe | — |
| $La_2Fe_2O_3Te_2$ | $La_2O_3 + 2FeTe$, annealing twice at 800°C | $La_2O_2Ch$, MO, MCh | — |
| $La_2Co_2O_3S_2$ | $La_2O_3 + 2CoCh$, annealing twice at 1000°C | | — |
| $La_2Co_2O_3Se_2$ | | $La_2Co_2O_3Se_2$, CoSe | $a = 4.0717(1)$, $c = 18.4288(4)$ [b)] |
| $Ce_2Co_2O_3Se_2$ | Ce + $CeO_2$ + CoSe, annealing twice at 1000° | | — |
| $La_2Ni_2O_3Se_2$ | $La_2O_3 + 2MSe$ | $La_2NiO_4$, ? | — |
| $La_2Zn_2O_3Se_2$ | | $La_2O_2Se$ + ZnO + ZnSe | — |

[a)] Ref. [7]: $a = 4.0263(1)$Å, $c = 18.4306(2)$Å.

[b)] Ref. [8]: $a = 4.0697(1)$Å, $c = 18.4198(2)$Å.

Table 2. Details of structural experiments for $Ln_2Fe_2O_3S_2$

| | | |
|---|---|---|
| Chemical formula | $Ce_2Fe_2O_3S_2$ | $Pr_2Fe_2O_3S_2$ |
| Crystal system | Tetragonal | |
| Space group | *I*4/*mmm* (# 139) | |
| Z | 2 | |
| Cell parameters: | | |
| *a*, Å | 4.0016(1) | 4.0093(1) |
| *c*, Å | 17.6524(6) | 17.6639(4) |
| V, $Å^3$ | 282.66(1) | 283.94(1) |
| Calculated density | 5.92(2) | 5.99(2) |
| Diffractometer, geometry | Stoe Stadi P, transmission | Bruker D8 Advance, reflection |
| $2\vartheta$ range, ° | 8-110 | 20-80 |
| # of reflections | 78 | 44 |
| # of structural parameters | 9 | 9 |
| Analyzing package | Topas [11] | |
| R values: | | |
| $R_B$ | 0.003 | 0.006 |
| $R_p$ | 0.013 | 0.017 |
| $R_{wp}$ | 0.017 | 0.022 |
| $\chi^2$ | 1.13 | 1.19 |

Table 3. Atomic parameters for $Ln_2Fe_2O_3S_2$

| Compound | $Ce_2Fe_2O_3S_2$ | | $Pr_2Fe_2O_3S_2$ | |
|---|---|---|---|---|
| Atom | *z* | *B*, Å$^2$ | *z* | *B*, Å$^2$ |
| Ln (0.5, 0.5, z) | 0.1809(1) | 0.85(5) | 0.1822(1) | 1.1(2) |
| Fe (0.5, 0, 0) | | 2.21(9) | | 0.8(2) |
| S (0, 0, z) | 0.0970(3) | 0.2(1) | 0.0960(5) | 0.4(2) |
| O1 (0.5, 0, 0.25) | | 1.0 | | 1.0 |
| O2 (0.5, 0.5, 0) | | 1.0 | | 1.0 |

Table 4. Bond distances and angles for $Ce_2Fe_2O_3S_2$

| Bond | Distance, Å | | Bonds | Angle, ° | |
|---|---|---|---|---|---|
| | Ln = Ce | Ln = Pr | | Ln = Ce | Ln = Pr |
| Ln – O1 | 2.343(1) | 2.335(1) | O1 – Ln – O1 | 105.72(4) × 4 | 105.27(4) × 4 |
| Ln – S | 3.194(2) | 3.218(3) | | 117.26(5) × 2 | 118.25(9) × 2 |
| Fe – O2 | 2.001(1) | 2.005(1) | S – Fe – S [a)] | 98.9(1) × 2 | 99.6(3) × 2 |
| Fe – S | 2.633(3) | 2.625(6) | | 81.1(1) × 2 | 80.4(3) × 2 |

[a)] Bond angles in distorted $TX_4$ squares in other $La_2Fe_2O_3Se_2$-type compounds: $La_2Fe_2O_3Se_2$: 97.0° and 83.0° [6], $Nd_2Fe_2O_3Se_2$: 96.1° and 83.9° [7], $La_2Co_2O_3Se_2$: 98.7° and 81.3° [8], $Sr_2Ti_2F_2OAs_2$: 96.3° and 83.7°; $Sr_2Ti_2F_2OSb_2$: 91.0° and 89.0° [12], $Sr_2Fe_2F_2OS_2$: 100.2° and 79.8°; $Ba_2Fe_2F_2OS_2$: 102.2° and 77.8°; $Sr_2Fe_2F_2OSe_2$: 97.2° and 82.8°; $Ba_2Fe_2F_2OSe_2$: 100.2° and 79.8° [10].

**Figures**

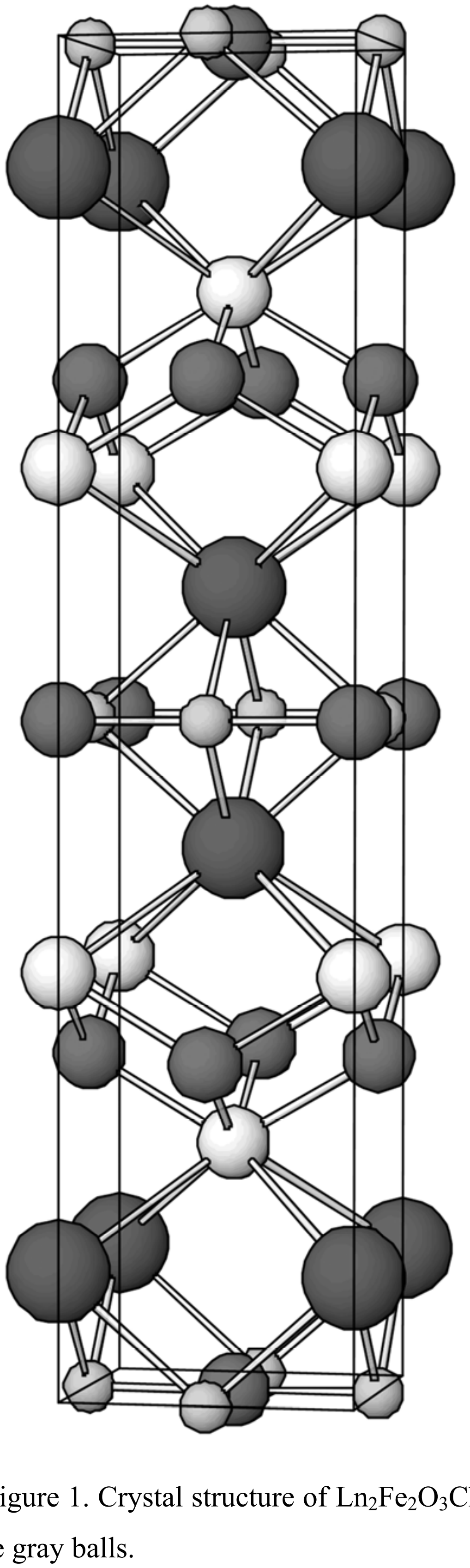

Figure 1. Crystal structure of $Ln_2Fe_2O_3Ch_2$. Ln: large white, Fe: small grayish, O: small gray, Ch: large gray balls.

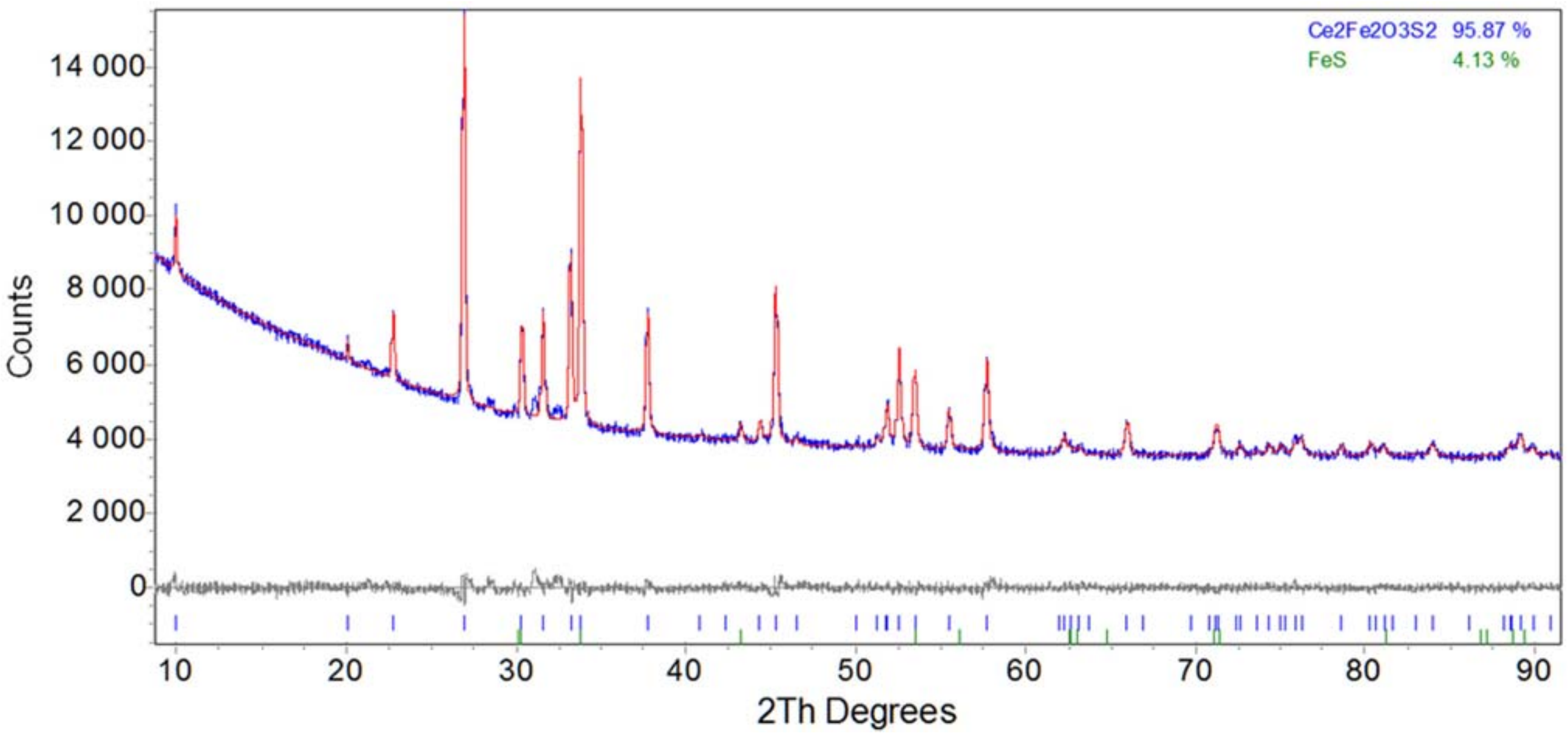

Figure 2. Final Rietveld refinement plot for $Ce_2Fe_2O_3S_2$.

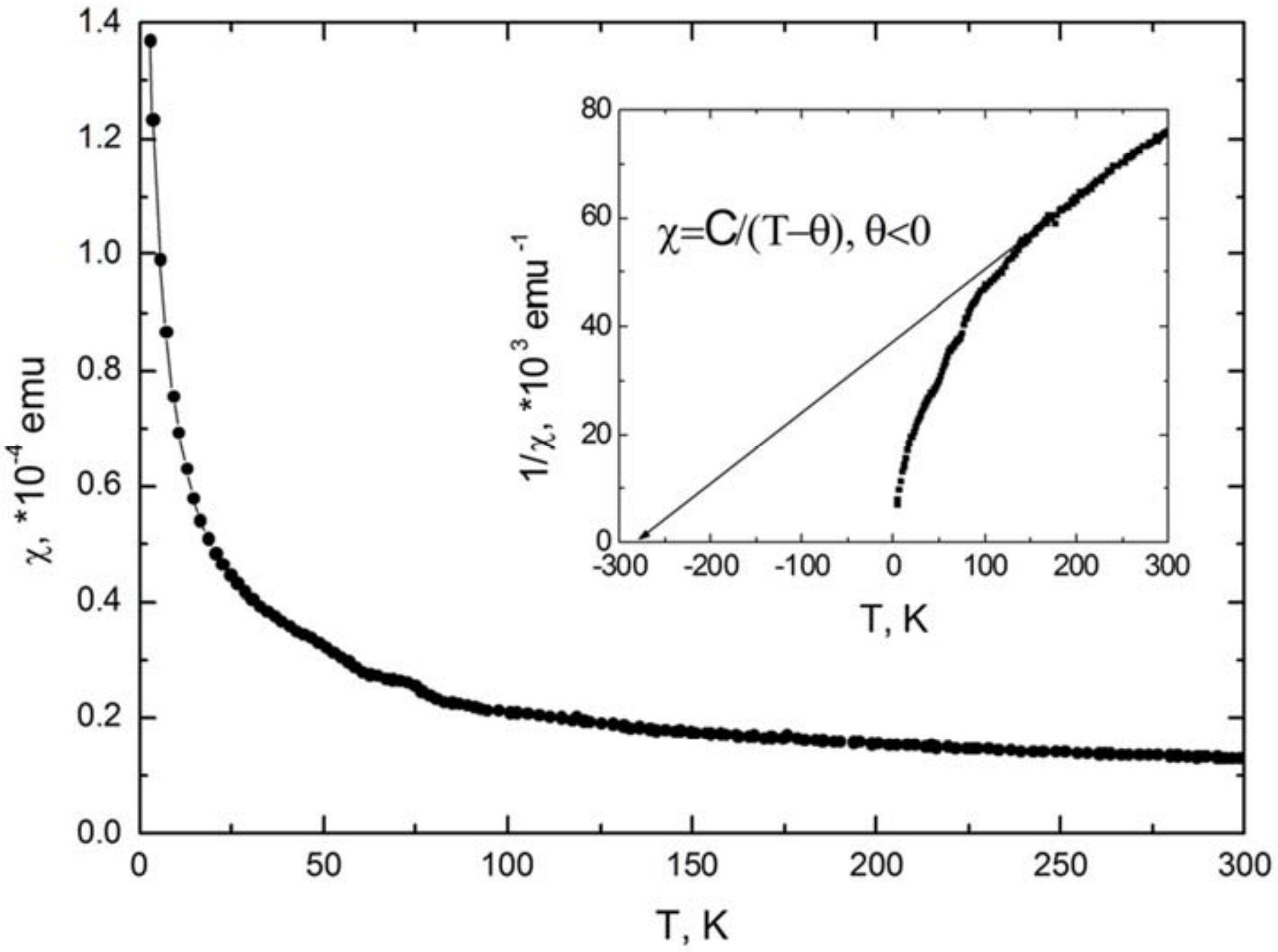

Figure 3. Temperature dependence of magnetic susceptibility of $Ce_2Fe_2O_3S_2$ measured in zero dc-field. The inset shows *anti*-ferromagnetic behavior at temperatures above 130K (Curie-Weiss law with $\theta<0$).